\documentclass{aa}     % LaTeX A&A New Fonts
\usepackage{epsfig}
\def\kms{km~s$^{-1}$ }
\def\a4{ABCG~496{}}
\def\bj{b$_{\rm J}${}}
\thesaurus{11.03.4 ABCG~85; 11.03.1 }

\title{A catalogue of velocities in the direction of the cluster of
galaxies Abell 496. 
\thanks{Based on observations collected at the European Southern
Observatory, La Silla, Chile}
\thanks{Tables 3 and 4 are only available in electronic form at the CDS via
anonymous ftp to cdsarc.u-strasbg.fr (130.79.128.5).}
}

 \author {
  F.~Durret \inst{1,2}
\and
  P.~Felenbok \inst{2}
\and
  C.~Lobo \inst{3}
\and
  E.~Slezak \inst{4}
}
\offprints{F.~Durret, durret@iap.fr }
\institute{
  Institut d'Astrophysique de Paris, CNRS, 
  98bis Bd Arago, F-75014 Paris, France 
\and 
    DAEC, Observatoire de Paris, Universit\'e Paris VII, CNRS (UA 173),
    F-92195 Meudon Cedex, France 
\and
    Osservatorio Astronomico di Brera, via Brera 28, I-20121 Milano, Italy
\and
    Observatoire de la C\^ote d'Azur, B.P. 229, F-06304 Nice Cedex 4, France 
}
\date{Received, 1999; accepted,}
\begin{document}

\maketitle

\begin{abstract}
We present a catalogue of velocities for 466 galaxies in the direction
of the cluster \a4 , in a region covering about 160'$\times$160'
(9.2$\times$9.2~Mpc for an average redshift for \a4 of 0.0331,
assuming H$_0$=50 km s$^{-1}$ Mpc$^{-1}$). This catalogue includes
previously published redshifts by Proust et al. (1987), Quintana \&
Ram\'\i rez (1990) and Malumuth et al. (1992), redshifts from the CfA
redshift survey, together with our new measurements.  A total of 274
galaxies have velocities in the 7800-11800~km~s$^{-1}$ interval, and
will be considered as members of the cluster. \a4 therefore becomes
one of the few clusters with a high number of measured redshifts; its
physical properties are investigated in a companion paper.\\
\end{abstract}

\section{Observations and data reduction} 

\subsection{Description of the observations}

The observations were performed with the ESO 3.6m telescope equiped
with MEFOS during 6 nights on November 5-11, 1994 and 2 nights on
November 24-26, 1995. A description of the instrument can be found in
Durret et al. (1998).  The grating used with the Boller \& Chivens
spectrograph had 300 grooves/mm, giving a dispersion of 224~\AA /mm in
the wavelength region 3820-6100~\AA . The detector was CCD \#32, with
512$^2$ pixels of $27 \times 27\ \mu$m.

The catalogue of galaxy positions used in this survey was obtained
with the MAMA measuring machine and is presented in a companion paper
(Slezak et al. 1999). This catalogue gives very accurate positions,
but magnitudes in the photographic \bj \ band are known with an
accuracy of 0.1 magnitude at best. CCD photometry of the central
regions of the cluster in the V and R bands was later performed to
recalibrate these \bj \ magnitudes and to obtain R magnitudes for the
entire photometric sample. We observed spectroscopically a total of 15
fields, with exposure times of 2$\times$30 minutes, and we obtained
473 spectra in total (plus the same number of sky spectra). The
limiting magnitude for the galaxies observed spectroscopically was \bj
=19, corresponding to a recalibrated magnitude R$\sim$18.7 (see
section~3).

\begin{figure}
\centerline{\psfig{figure=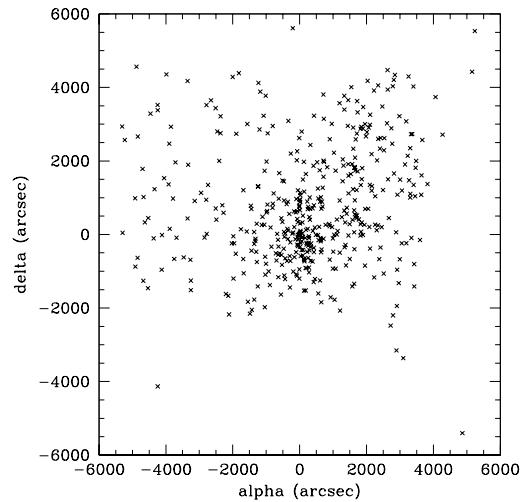,height=7cm}}
\caption[ ]{Spatial distribution of the 466 galaxies with redshifts in the
direction of ABCG 496 (map limited to $\pm$6000 arcsec from the cluster 
center).}
\protect\label{xy}
\end{figure}

Out of the 473 spectra obtained, we measured 410 reliable redshifts
(the other ones were discarded due to insufficient signal to noise
ratio). Since \a4 is at a relatively low galactic latitude, the
effects of stellar contamination of our photographic plate catalogue
are large. Our star-galaxy separation software was imperfect since out
of the 410 objects for which redshufts were measured only 305 were
galaxies.  Such a high degree of contamination of the catalogue is at
least partly due to the fact that we preferred to have a galaxy
catalogue as complete as possible, in order not to ``miss'' galaxies.
Our catalogue includes these 305 galaxy spectra, plus those previously
published by Proust et al. (1987), Quintana \& Ram\'\i rez (1990) and
Malumuth et al. (1992), and some galaxies from the CfA redshift
catalogue (Huchra et al. 1992), reaching a total of 466 galaxy
redshifts after eliminating objects observed twice.

The positions of the objects for which we gathered reliable spectra
(either from our observations or from the literature) are shown in
Fig.~\ref{xy}. These positions are relative to the cluster center
taken to be the position of the maximum X-ray emission (Pislar 1998):
$\alpha_{2000.0} = 04^h33^{m}37.9^s, \delta_{2000.0} = -13^\circ
15'47''$. This center is within 7 arcsec of the position of the cD
galaxy, a distance which is smaller than the ROSAT PSPC pixel size,
and we will therefore consider hereafter that both positions coincide.

\subsection{Data reduction}

The spectra were reduced using the IRAF software. The frames were bias
corrected in the usual way. Velocities were measured by
cross-correlating the observed spectra with different templates: a
spectrum of M31 (kindly provided by J. Perea) at a velocity of
$-300$~km~s$^{-1}$, and stellar spectra of the standard stars HD~24331
and HD~48381, which were each observed every night during our 1994
run.  The cross-correlation technique is that described by Tonry \&
Davis (1979) and implemented in the XCSAO task of the RVSAO package in
IRAF (Kurtz et al.  1991). The errors on the velocities derived from
absorption lines are given automatically by this task.

The positions of emission lines, when present, were measured by
fitting each line with a gaussian.

All the redshifts were measured by the same person (F.D.) in a
homogeneous way.  Redshifts of insufficient quality were discarded
(i.e. those with a Tonry \& Davis parameter smaller than 2.0, except
for three galaxies with respective Tonry \& Davis parameters of 1.5,
1.6 and 1.9 where the absorption lines seemed to be well enough
defined for these redshifts to be kept in the final catalogue).

\section{Quality of the data}

\begin{table*}
\caption{Comparison of galaxy velocities measured by us to those in the 
literature: QR = Quintana \& Ram\'\i rez 1990, M = Malumuth et al. 1992) }
\begin{tabular}{rrrrrrrrr}
\hline
Coordinates & QR & QR & QR & Our & Our & Our & Velocity & R\\
(2000.0)    & velocity  & error    & number& velocity & error &reference & 
difference & magnitude\\
\hline
4 31  3.14 -12 58 27.01 & 10457 & 33 & QR32 & 10524 & 22 &  73 & 67 & 14.4 \\ 
4 34  9.86 -13 14 35.63 & 10044 & 38 & QR358&  9932 & 27 & 331 & -112 & 16.6\\
\hline
Coordinates & M & M & M & Our & Our & Our & Velocity & R \\
(2000.0)    & velocity  & error    & number& velocity & error &reference & difference & magnitude \\
\hline
4 31 23.63 -13 04 12.39 & 9797 & 126 &  M3 & 9610 & 76 &  90 & -187 & 15.5 \\
4 31 43.06 -13 07 09.70 &10991 & 163 &  M5 &11106 & 43 & 115 & 115 & 17.0 \\ 
4 32 23.93 -12 46 47.45 &10700 &  30 & M19 &10140 & 23 & 164 & -560 & 14.9 \\ %*
4 32 52.84 -13 22 33.42 & 9665 &  80 & M27 & 9594 & 21 & 189 & -71 & 15.7 \\ 
4 33 00.06 -13 15 59.91 &11063 &  51 & M39 &10967 & 19 & 204 & -96 & 15.7 \\ 
4 33 11.43 -13 27 59.68 &10689 & 170 & M50 &10670 & 19 & 218 & -19 & 17.5 \\  %*
4 33 14.38 -13 19 48.88 &11802 &  74 & M51 &11860 & 27 & 219 & 58 & 17.3 \\  
4 33 28.34 -13 27 18.83 & 9806 & 144 & M74 &10134 & 49 & 249 & 328 & 18.1 \\ 
4 33 33.49 -13 18 52.17 &11773 &  56 & M83 &11754 & 31 & 262 & -19 & 17.5 \\ 
4 33 33.92 -13 22 50.20 &10235 &  45 & M84 &10288 & 20 & 264 & 53 & 16.5 \\  
4 33 38.40 -13 02 39.49 &16691 & 136 & M93 &16808 & 50 & 281 & 117& 18.3 \\  
4 33 42.78 -13 08 46.79 &10495 &  50 &M102 &10546 & 58 & 295 & 51 & 17.9 \\   %*
4 33 56.88 -13 19 18.75 & 9561 &  77 &M114 & 9577 & 41 & 313 & 16 & 17.7 \\  
4 34 03.81 -13 27 47.23 & 8305 &  42 &M121 & 8316 & 58 & 323 & 11 & 15.8 \\  
4 34 07.25 -13 15 14.45 & 9440 &  51 &M123 & 9521 & 22 & 326 & 81 & 16.6 \\  
4 34 12.95 -13 10 03.83 &10836 &  64 &M128 &10894 & 58 & 335 & 58 & 17.3 \\  
4 34 35.72 -13 21 19.38 &10093 &  27 &M139 &10100 & 23 & 354 &  7 & 17.0 \\  
4 35 32.37 -13 33 22.23 & 9867 & 112 &M154 &10121 & 23 & 395 & 254 & 15.5 \\ 
\hline
Coordinates & CfA & CfA & CfA & Our & Our & Our & Velocity & R \\
(2000.0)    & velocity  & error    & number& velocity & error &reference & difference & magnitude \\
\hline
4 31 03.22 -12 58 26.12 &10519 &  33 & A0428-1304 &10524 & 22 &  73 &   -5 & 14.4 \\ %*
4 31 16.57 -12 27 18.90 & 9582 &  32 &      I 376 & 9224 & 23 &  85 & -358 & 14.0 \\ %* 
4 31 23.71 -13 04 13.44 & 9797 & 126 & A0429-1310 & 9610 & 76 &  90 & -187 & 15.5 \\ 
4 31 43.10 -13 07 10.74 &10991 & 163 & A0430-1253A&11106 & 43 & 115 &  115 & 17.0 \\  %*
4 32 52.92 -13 22 33.43 & 9581 &  28 & A0430-1328A& 9594 & 21 & 189 &   13 & 15.7 \\ 
4 33 00.14 -13 16 00.08 &11062 &  34 & A0430-1322B&10967 & 19 & 204 &  -95 & 15.7 \\ 
4 33 14.43 -13 19 49.28 &11802 &  74 & A0430-1326 &11860 & 27 & 219 &   58 & 17.3 \\ 
4 33 28.41 -13 27 19.68 & 9806 & 144 & A0431-1333A&10134 & 49 & 249 &  328 & 18.1 \\ 
4 33 33.53 -13 18 52.51 &11773 &  56 & A0431-1325C&11754 & 31 & 262 &  -19 & 17.5 \\ 
4 33 33.92 -13 22 50.71 &10347 &  34 & A0431-1329C&10288 & 20 & 264 &  -59 & 16.5 \\ 
4 33 38.46 -13 02 39.65 &16691 & 136 & A0431-1308A&16808 & 50 & 281 &  117 & 18.3 \\ 
4 33 56.93 -13 19 19.48 & 9561 &  77 & A0431-1325E& 9577 & 41 & 313 &   16 & 17.7 \\ 
4 34 07.24 -13 15 14.35 & 9513 &  47 & A0431-1321H& 9521 & 22 & 326 &    8 & 16.6 \\ 
4 34 09.94 -13 14 35.07 &10101 &  38 & A0431-1320D& 9932 & 27 & 331 & -169 & 16.6 \\  %*
4 34 13.05 -13 10 04.09 &10836 &  64 & A0431-1316C&10894 & 58 & 335 &   58 & 17.3 \\ 
4 34 35.72 -13 21 19.45 &10191 &  29 & A0432-1327B&10100 & 23 & 354 &  -91 & 17.0 \\ 
4 35 32.40 -13 33 22.54 & 9867 & 112 & A0433-1339 &10121 & 23 & 395 &  254 & 15.5 \\ 
\hline
\end{tabular}
\protect\label{doubles}
\end{table*}

\begin{figure}
\centerline{\psfig{figure=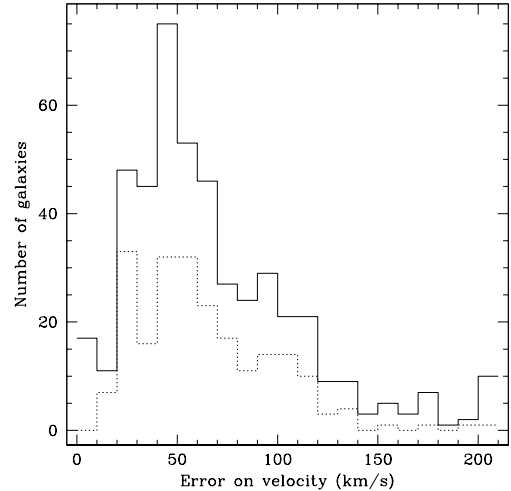,height=7cm}}
\caption[ ]{Distribution of errors on velocities derived from
absorption lines for the entire catalogue (full line) and for our data
(dotted line). }
\protect\label{errorvabs}
\end{figure}

\begin{figure}
\centerline{\psfig{figure=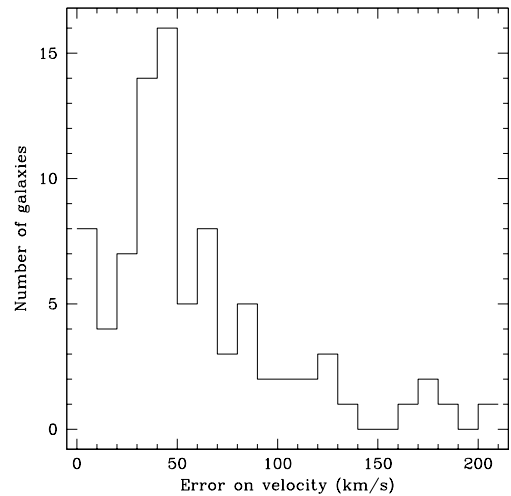,height=7cm}}
\caption[ ]{Distribution of errors on velocities derived from 
emission lines (our data only).}
\protect\label{errorvem}
\end{figure}

\begin{figure}
\centerline{\psfig{figure=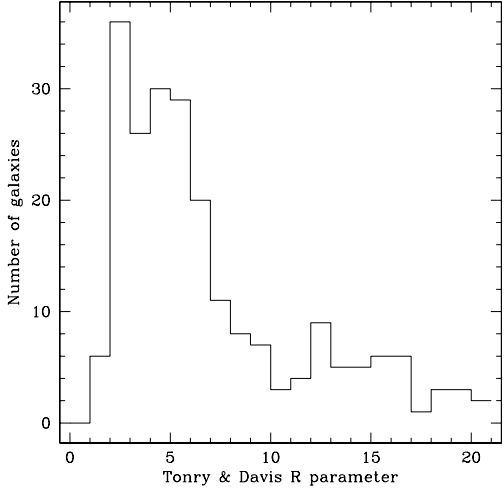,height=7cm}}
\caption[ ]{Distribution of the Tonry \& Davis R parameter given by the 
cross-correlation measure on absorption lines (our data).}
\protect\label{Rparam}
\end{figure}

\begin{figure}
\centerline{\psfig{figure=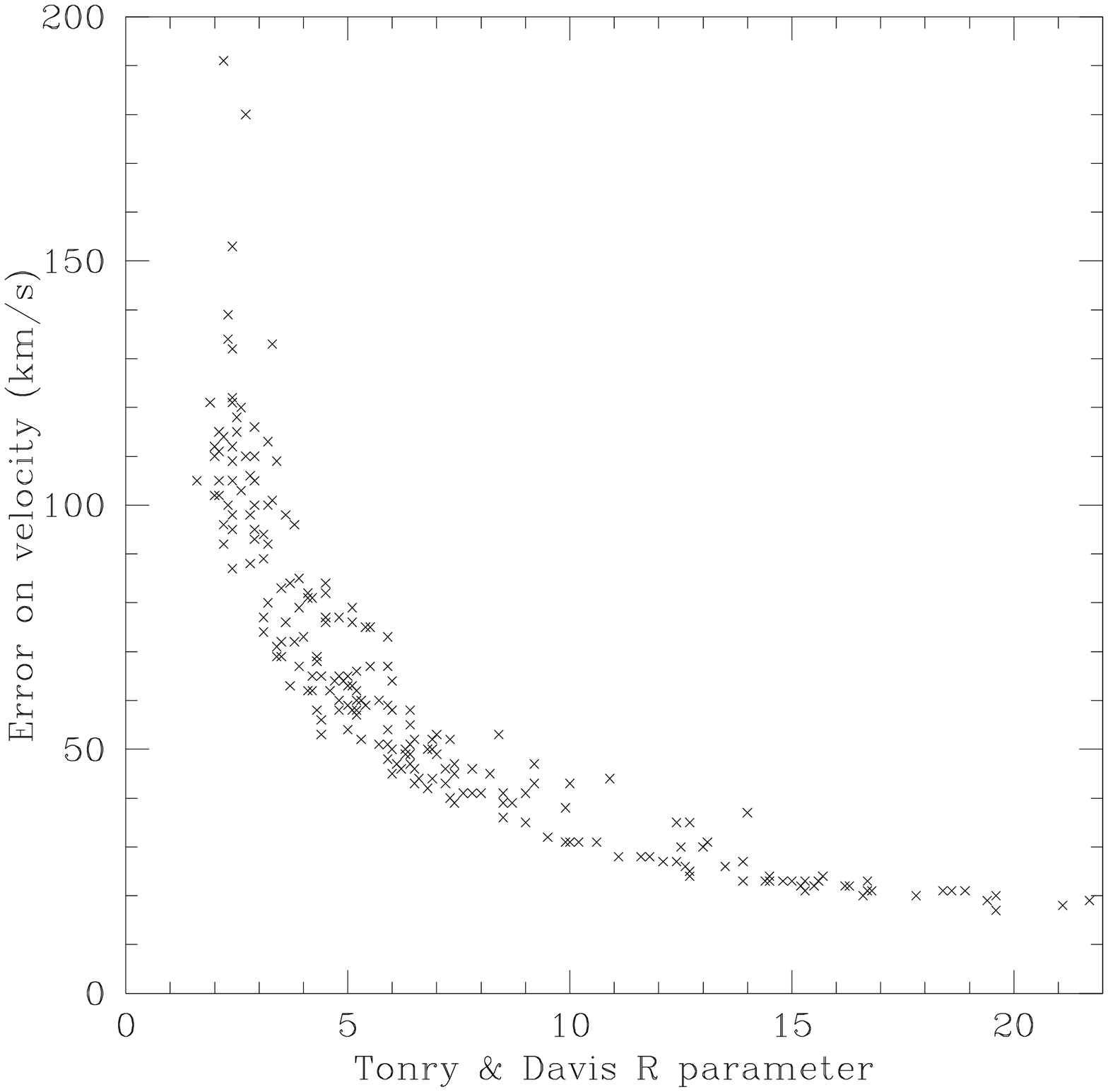,height=7cm}}
\caption[ ]{Relation between the Tonry \& Davis R parameter and the
errors on velocities derived from absorption lines (our data).}
\protect\label{RTDerr}
\end{figure}

For galaxies with absorption lines, two velocity standard stars from
the Maurice et al. (1984) list were observed each night in order to
check the intrinsic quality of our velocity measurements. The errors,
derived by cross-correlating the star spectra to the spectrum of M31,
range (from night to night) from $\pm$16 to $\pm 23$~km~s$^{-1}$ for
HD~24331, and from $\pm$17 to $\pm 43$~km~s$^{-1}$ for HD~48381. The
mean internal error on velocities derived from the {\sl rms} in the
mean wavelength calibration is 66 km~s$^{-1}$.

For emission line measurements, the errors on velocities were
estimated from the dispersion of the velocities derived from the
various emission lines present. When only one emission line was
present we averaged the emission and absorption line redshifts
whenever possible; if no reliable absorption line redshift was
available, we estimated the internal error on a single emission line
to be the intrinsic value of 66~km~s$^{-1}$. The number of redshifts
obtained from emission lines is 85.

The distributions of errors on all the velocities in the catalogue are 
displayed in Figs. \ref{errorvabs} and \ref{errorvem} for absorption
and emission line measurements respectively.  For the 220 galaxies
with absorption lines taken from our observations, the histogram of
the Tonry \& Davis signal to noise parameter R given by the
cross-correlation measure is displayed in Fig.~\ref{Rparam} (this
quantity is not given in previously published catalogues). The
corresponding correlation between the Tonry \& Davis R parameter and
the error on the velocity is shown in Fig.~\ref{RTDerr}.

In order to check the quality of our redshifts, we also reobserved 2
galaxies from Quintana \& Ram\'\i rez (1990) and 18 from Malumuth et
al. (1992). The results are shown in Table \ref{doubles}. For galaxies
observed twice, we chose to give in our final catalogue the redshift
with the smallest error (usually our data). The mean absolute
difference between our measurements and those of Malumuth et al. (18
galaxies) is 117 \kms, with a dispersion of 137 \kms, implying that
the general agreement is good. Note that the difference between our
velocities and those of the literature does not tend to be larger for
fainter magnitudes.  The agreement with the two galaxies in common
with Quintana \& Ram\'\i rez (1990) is satisfactory but cannot be
tested statistically. A comparison with 17 galaxies in common with the
CfA redshift catalogue (Huchra et al. 1992) gives a mean absolute
difference of 115 \kms with a dispersion of 107 km s$^{-1}$. Notice
that there are many galaxies in common in the Malumuth and CfA
samples.

The final redshifts derived from our observations given in the
catalogue are those derived from the cross-correlation with M31, since
this template gave the best results.  A correction was applied to
correct for the velocity of M31 ($-$300 \kms) and to obtain heliocentric
velocities.

\begin{figure}
\centerline{\psfig{figure=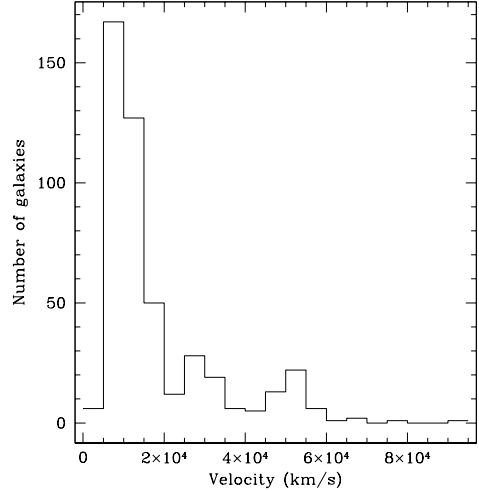,height=7cm}}
\caption[ ]{Velocity histogram of all the galaxies in the catalogue.}
\protect\label{histov}
\end{figure}

The histogram of all the velocities in the catalogue is displayed in 
Fig.~\ref{histov}. It will be discussed in detail in a companion paper
(Durret et al. in preparation).

\begin{table*}
\caption{Completeness of the redshift catalogue for different field
diameters and for various limiting magnitudes in the R band.  Numbers
in parentheses indicate the absolute numbers of galaxies with and
without redshifts respectively. Note that for galaxies belonging to
the cluster 1800'' correspond to 1.732~Mpc
with H$_\circ$=50~km~s$^{-1}$~Mpc$^{-1}$.}

\begin{tabular}{llllll}
\hline
              &       &      &      & \\
Limiting      & 1800  & 2400 & 3600 & 5400 \\
diameter ('') &       &      &      & \\
R magnitude   &       &      &      & \\
              &       &      &      & \\
\hline
          &  &      & & & \\
16.0 &100.0\%~(64/64)   & 98.7\%~(77/78)   & 80.7\% (96/119)  & 62.3\%~(114/183) \\
16.5 &~84.3\%~(75/89)   & 74.6\%~(94/126)  & 65.1\% (123/189) & 52.1\%~(146/280) \\
17.0 &~81.7\%~(98/120)  & 72.3\% (128/177) & 61.4\% (178/290) & 51.1\%~(216/423) \\
17.5 &~85.5\% (124/145) & 74.0\% (165/223) & 62.0\% (230/371) & 50.4\%~(289/573) \\
18.0 &~80.6\% (158/196) & 70.5\% (206/292) & 57.3\% (283/494) & 46.0\%~(354/769) \\
18.5 &~78.7\% (196/249) & 69.1\% (250/362) & 52.7\% (336/638) & 40.5\%~(415/1024) \\
19.0 &~62.6\% (206/329) & 55.5\% (263/474) & 40.6\% (353/869) & 30.6\% (435/1421) \\
          &  &      & & & \\
\hline
\end{tabular}
\protect\label{compl}
\end{table*} 

We have estimated the completeness of the final spectroscopic
catalogue presented here by comparing the number of galaxies with
redshifts to the total number of galaxies from our photographic plate
catalogue in the same area. Results are shown in
Table~\ref{compl}. Note that the 17 galaxies in our redshift catalogue
which do not have magnitudes (see section 3) have been excluded from
these statistics, and therefore the catalogue completeness may be
slightly larger than estimated in the Table.

\section{Galaxy magnitudes}

\begin{figure}
\centerline{\psfig{figure=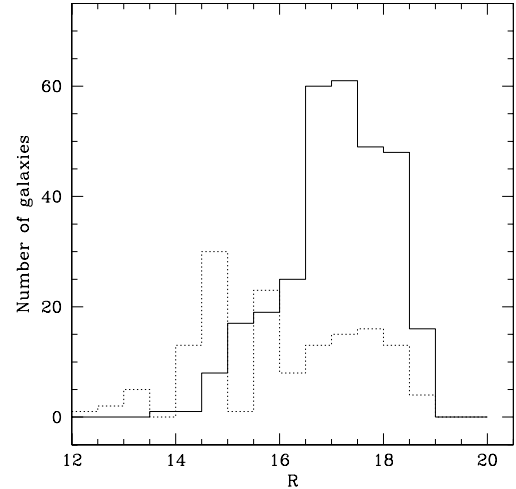,height=7cm}}
\caption[ ]{Magnitude histogram in the R band for the 305 galaxies 
for which we measured velocities (full line) and for the 144 galaxy redshifts
taken from the literature (dotted line). The 17 missing galaxies are those
for which no magnitude is available.}
\protect\label{maggal}
\end{figure}

\begin{figure}
\centerline{\psfig{figure=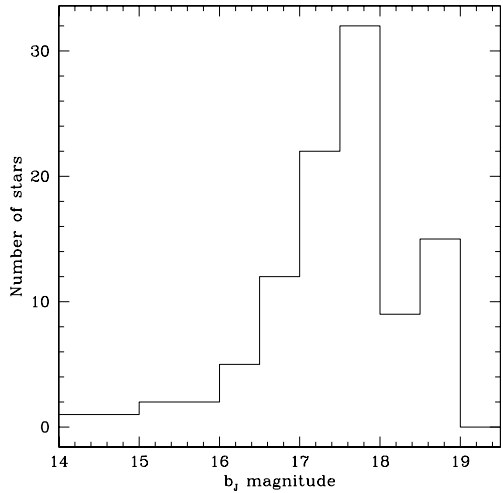,height=7cm}}
\caption[ ]{Magnitude histogram in the \bj\ band of the 101 stars
initially misclassified as galaxies in the photometric sample used
for our target selection.} 
\protect\label{magstars}
\end{figure}

The magnitude histograms for the galaxies and stars misclassified as
non-PSF-like objects are displayed in Figs.~\ref{maggal} and
\ref{magstars} respectively.  The percentage of contamination by stars
is very high: about 20\%.

For a large fraction of the galaxies taken from the literature, R
magnitudes were made available to us by C. Adami (private
communication). In order to give magnitudes for all galaxies in the
same photometric band, we estimated the R band magnitudes for the
galaxies that we observed. This was done by identifying galaxies both
in our photographic plate catalogue and in our CCD catalogue, and by
finding the best linear fit (see Slezak et al. 1999 for details). This
best fit was: $ {\rm R}=b_{\rm J} -0.28 $, with the slope fixed to
1.0, giving an {\sl rms} fluctuation of 0.04. The galaxies in the
literature for which Adami did not have magnitudes were identified
with objects in our photographic plate catalogue, and we applied the
same relation to derive their R magnitudes from their b$_{\rm J}$
magnitudes. For some of the CfA galaxies, only B$_{\rm T}$ magnitudes
were available; we roughly transformed them to R magnitudes, by
calculating the mean (V$-$R) colour for the 239 galaxies detected in V
in our CCD catalogue (Slezak et al. 1999); this mean value is equal to
0.48, and therefore roughly corresponds to a G0 star for which
(V$-$R)=0.52 and (B$-$V)=0.58 (e.g. Allen 1981), leading to
R$\sim$B$-$1.1; for these CfA galaxies, we therefore took R= B$_{\rm
T} -1.1$.

33 galaxies in our redshift catalogue have no identification with our
photometric catalogue: 4 from Quintana \& Ram\'\i rez (1990), 14 from
Malumuth et al. (1992), 1 from Proust et al. (1987) and 14 from Huchra
et al. (1992). They are the objects in Table 4 that have labels 2 to 5
in column (16), and their corresponding identification number in the
literature, but no identification number from our photographic plate
catalogue. Note that out of these 33 objects, half are located outside
the area covered by our photometric catalogue, and three are located
less than 1 arcmin from the cluster center, so we cannot separate them
from the cD.  The reason for the lack of identification of the other
objects is not clear, since in all the cases where we did identify
galaxies from our sample to galaxies in the literature the coordinates
matched within a few arcseconds. The most likely explanation is that
the coordinates given in the literature for these objects are not
accurate; the fact that they may have not been detected in our
photographic plate scan is unlikely since they are all very bright
(R$\leq$16.8).

\section{The catalogues}

The coordinates and \bj magnitudes of the 101 stars misclassified as
galaxies are given in Table 3 (available in electronic form only), to
avoid further observations of these objects in galaxy surveys.

The velocity data for the galaxies in the field of ABCG~496 are given 
in Table~4 (available in electronic form only). The meaning of the 
columns is the following:\\
(1) running number; \\
(2) to (4) right ascension (equinox 2000.0); \\
(5) to (7) declination (equinox 2000.0); \\
(8) heliocentric velocity (cz) in \kms; \\
(9) error on the velocity in \kms; for velocities derived from absorption 
lines, the error is either that stated in the literature or, for our own 
measurements, that given by the RVSAO IRAF package;  for velocities derived 
from several emission lines, the error was estimated from the dispersion
between the velocities derived from the different emission lines; 
when only one emission line was present and no absorption line redshift was 
available, the error on the velocity was taken to be the mean internal 
velocity error; \\
(10) R band magnitude (see section 3);\\
(11) and (12) X and Y positions in arcseconds relative to the center
assumed to have coordinates: \\
$\alpha_{2000.0} = 04^h33^{mn}37.9^s, \delta_{2000.0} = -13^\circ 15'47''$;\\
(13) distance to the cluster center in arcseconds;\\
(14) Tonry \& Davis parameter;\\
(15) label indicating the means of determination of the redshift: 0=derived 
from absorption lines, 1=derived from emission lines; \\
(16) label indicating the origin of the data: 1=our data,
2=Quintana \& Ram\'\i rez (1990), 3=Malumuth et al. (1992),
4=Proust et al. (1987), 5=Huchra et al. (1992); 
note that the catalogue by Malumuth et al. (1992)
includes previous measurements by Quintana et al. (1985) and Zabludoff
et al. (1990);\\
(17) reference to the galaxy number in the various catalogues; for
our own data (label 1 in col. 16), the number appearing in this column
is that of the photographic plate catalogue by Slezak et al. (1999),
Table 1; for data taken from the literature, the first number is that
of the previously published catalogue (or the name for CfA galaxies),
the second that of the photographic plate catalogue by Slezak et
al. (1999) whenever the identification was possible.\\

\section{Conclusions} 

This redshift catalogue will be used, together with our photometric
catalogues (which include both the large field catalogue obtained by
scanning a photographic plate and the small CCD field catalogue,
Slezak et al. 1999), to give an interpretation of the properties of
\a4 (Durret et al. in preparation).  They will be compared to the
results already obtained from ROSAT PSPC X-ray data (Pislar 1998).

\acknowledgements {We are very grateful to Andr\'ee Fernandez for her
help during the preparation of the first observing run, to Paul Stein
for helping us obtain a catalogue of guiding stars and to Cl\'audia
Mendes de Oliveira for her cheerful and highly competent assistance at
the telescope. We also wish to thank the referee, John Huchra, for
providing us with the redshift list from the CfA redshift catalogue in
the direction of this cluster, and warmly acknowledge the help of
Fran\c coise Warin in the data reduction. Last but not least, we are
very grateful to Christophe Adami and Andrea Biviano for giving us the
catalogue of velocities and R magnitudes taken from the literature in
electronic form. C.L. acknowledges financial support by the CNAA
(Italia) fellowship reference D.D. n.37 08/10/1997. }

\end{document}